\documentclass[pra,aps,amsmath,amssymb,showpacs,twocolumn]{revtex4}

\usepackage{graphicx}
\usepackage{amsmath}
\newcommand{\Hil}{{\mathcal H}}
\newcommand{\Ell}{{\mathcal L}}

\newcommand{\Boltz}{ k_{\rm\scriptscriptstyle B}}

\newcommand{\Tr}{{\rm Tr}}

\newcommand{\ddt}[1]{{\frac{\displaystyle {\rm d}#1}{\displaystyle {\rm d}t}}}

\newcommand{\sq}{{\sqrt{\rho}}}
\newcommand{\sqnd}{{\sqrt{\rho_{\rm nd}}}}

\newcommand{\cov}[2]{{\langle\Delta #1\Delta #2\rangle}}
\newcommand{\sqcov}[2]{{\sqrt{\sigma_{\!#1\!#2}}}}
\newcommand{\scov}[2]{{\sigma_{\!#1\!#2}}}
\newcommand{\sccov}[2]{{\sigma^2_{\!#1\!#2}}}
\newcommand{\ceta}[2]{{\eta_{#1\!#2}}}
\newcommand{\cseta}[2]{{\eta^*_{#1\!#2}}}
\newcommand{\cceta}[2]{{\eta^2_{#1\!#2}}}
\newcommand{\com}[2]{{\langle [ #1, #2]/2i\rangle}}
\newcommand{\mean}[1]{{\langle #1\rangle}}

\newcommand{\citen}[1]{{ \cite{#1}}}

\newcommand{\N}{{\mathbf N}}
\newcommand{\mmu}{{\boldsymbol{\mu}}}

\newcommand{\Pen}{{P_{\!e_{\!n}}}}
\newcommand{\Pem}{{P_{\!e_{\!m}}}}

\newcommand{\Petwo}{{P_{\!e_{2}}}}
\newcommand{\Pethree}{{P_{\!e_{3}}}}

\newcommand{\rFM}{{r_{\!F\!M}}}
\newcommand{\cFH}{{c_{F\!H}}}
\newcommand{\cMH}{{c_{\!M\!H}}}
\newcommand{\cPenH}{{c_{\Pen\!\!H}}}
\newcommand{\rPenM}{{r_{\Pen\!\!M}}}
\newcommand{\rAM}{{r_{\!A\!M}}}
\newcommand{\rSM}{{r_{\!S\!M}}}

\newcommand{\rSHprime}{{r_{\!S\!H'}}}
\newcommand{\rrFH}{{r^2_{\!F\!H}}}

\newcommand{\ccFH}{{c^2_{\!F\!H}}}
\newcommand{\rrSM}{{r^2_{\!S\!M}}}

\newcommand{\rrSHprime}{{r^2_{\!S\!H'}}}
\newcommand{\rrAM}{{r^2_{\!A\!M}}}
\newcommand{\PRA}{{Phys. Rev. A\ }}
\newcommand{\PRD}{{Phys. Rev. D\ }}
\newcommand{\PRE}{{Phys. Rev. E\ }}
\newcommand{\PRL}{{Phys. Rev. Lett.\ }}

\newcommand{\JPA}{{J. Phys. A\ }}

\newcommand{\FoundPhys}{{Found. Phys.\ }}
\newcommand{\NCB}{{Nuovo Cimento Soc. Ital. Fis., B\ }}

\newcommand{\JMP}{{J. Math. Phys.\ }}
\newcommand{\EPJB}{{Eur. Phys. J.\ B\ }}
\newcommand{\MPLA}{{Mod. Phys. Lett.\ A\ }}
\newcommand{\PLA}{{Phys. Lett.\ A\ }}
\newcommand{\be}{\begin{equation}}
\newcommand{\ee}{\end{equation}}
\newcommand{\bea}{\begin{eqnarray}}
\newcommand{\eea}{\end{eqnarray}}
\newcommand{\fr}{\frac}

\begin{document}

\title{Time-energy and time-entropy  uncertainty relations in
 dissipative quantum
dynamics}
\author{ Gian Paolo Beretta }
\affiliation{ Universit\`a di Brescia, via Branze 38, 25123
Brescia, Italy } \email{beretta@unibs.it}
\date{\today}

\begin{abstract} We derive exact relations and general
 inequalities that extend the usual time-energy uncertainty relations
  from the domain of unitary Hamiltonian dynamics to that of dissipative
  dynamics as described by a broad class of linear and nonlinear evolution
   equations for the density operator.
   We restrict our attention to intrinsic characteristic times of dynamical
   variables associated with the linear functionals $\Tr(\rho F)$ of the
   density operator, as well as with the nonlinear entropy functional
   $-\Boltz\Tr(\rho\ln\rho)$.  For non-dissipative
    dynamics, by using the Schroedinger inequality instead
   of the Heisenberg-Robertson inequality, we obtain a general exact
    time-energy uncertainty relation which is sharper than the usual
    Mandelstam-Tamm-Messiah relation
    $\tau_{F}\Delta_H\ge \hbar/2$. For simultaneous
    unitary/dissipative dynamics, the usual
    time-energy uncertainty relation is
   replaced
   by a less restrictive relation that depends on the characteristic
    time of dissipation, $\tau$, and the uncertainty associated with
    the generalized nonequilibrium Massieu-function operator which defines the
    structure of the dissipative part
    of the assumed class of evolution equations.  Within the steepest-entropy-ascent
    dissipative quantum
    dynamics of an isolated system introduced earlier by this author,
    we obtain the interesting time-energy
    and time-entropy uncertainty relation $(2\tau_{F}\Delta_H/ \hbar)^2+
(\tau_{F}\Delta_S/\Boltz\tau)^2 \ge 1$. We illustrate this result
and various
    other inequalities
    by means of numerical simulations. \end{abstract}


\maketitle


\section{Introduction}

The time-energy uncertainty relation  has remained an open and at
times controversial issue throughout the history of quantum
theory. Several reviews are available on the pioneering
discussions  and the more recent developments \cite{reviewsold}.
 We are motivated by the recent revival of
foundational questions and the development of dynamical theories
that seek to build into the laws of quantum mechanics those of
equilibrium and non-equilibrium thermodynamics. Such revival is
currently paralleled by the steady advancement of experimental
techniques dealing with single ion traps \cite{iontraps}, qubits
\cite{qubits}, neutron interferometry \cite{neutron}, and a
growing number of other developments all pointing at microscopic
few-particle setups that nevertheless exhibit non-unitary
dissipative dynamical behaviour and call for the investigation of
the role of general thermodynamic principles at the microscopic
quantum level. Here we address the question of how the usual
time-energy uncertainty relation, as interpreted according to the
Mandelstam-Tamm-Messiah intrinsic-time approach \cite{Messiah}
based on unitary Hamiltonian dynamics, is modified by the presence
of dissipation in the dynamical law or model.

Let $\Hil$ (${\rm dim}\Hil \le \infty$) be the Hilbert space and
$H$ the Hamiltonian operator  associated with a system in standard
Quantum Mechanics.   We assume that the  quantum states are
one-to-one with the linear hermitian operators $\rho$ on $\Hil$
with $\Tr(\rho)=1$ and $\rho\ge \rho^2$, and we assume a dynamical
equation of the form
\begin{equation}
\label{rhodot}\ddt{\rho}=\rho\,E(\rho)+ E^\dagger(\rho)\,\rho \ ,
\end{equation}
where $E(\rho)$ is an operator-valued function of $\rho$ that we
may call
 the
``evolution" operator and in general is non-hermitian.  Without
loss of generality, we  write  $E=E_+ +iE_-$ where
$E_+=(E+E^\dagger)/2$ and $E_-=(E-E^\dagger)/2i$ are hermitian
operators that, for convenience, we rename as  $\Delta
M(\rho)/2\Boltz\tau(\rho)$ and $H/\hbar$, respectively, so that
Eq.\ (\ref{rhodot}) takes the form
\begin{equation}
\label{rhodotHM}\ddt{\rho}=-\frac{i}{\hbar}[H,\rho]+
\frac{1}{2\Boltz\tau(\rho)}\{\Delta M(\rho),\rho\} \ ,
\end{equation}
where $[\,\cdot\,,\,\cdot\,]$ and  $\{\,\cdot\,,\,\cdot\,\}$ are
the usual commutator and anticommutator, $H$ (assumed independent
of $\rho$, but not necessarily independent of time $t$) is
identified with the Hamiltonian operator, $\hbar$ the reduced
Planck constant, $\Boltz$ the Boltzmann constant; moreover,
$\Delta M(\rho)$ is a hermitian operator-valued, possibly
nonlinear function of $\rho$ which, together with the positive
definite, possibly nonlinear functional $\tau(\rho)$ of $\rho$,
describes the dissipative dynamics of the system, and is such that
$\Tr[\rho\Delta M(\rho)]=0$ as required to preserve $\rho$ unit
trace.

The reason for considering a dynamical law of the form
(\ref{rhodotHM}) is that the explicit expression of $\Delta
M(\rho)$ that generates steepest-entropy-ascent (maximal entropy
generation) conservative dynamics of an isolated system compatible
with all thermodynamics requirements \cite{MPLA} is known
\cite{Beretta}. We  use it in Sections \ref{Example} and
\ref{Numerical} as an illustration of our general results, within
a model for irreversible relaxation in a Boltzmann gas of
$N$-level atoms. The extension to structured composite systems,
which is nontrivial in view of the assumed nonlinearity of the
dynamical law with respect to $\rho$, will be discussed elsewhere.
Additional discussion about the form of equation (\ref{rhodotHM})
is given in the Appendix.

The steepest-entropy-ascent form of the operator function $\Delta
M(\rho)$ is discussed in Section \ref{Example}, after introducing
the necessary notation,  in terms of the operator
\begin{equation}
\label{Massieu} M(\rho)=  S(\rho)-\frac{ H}{\theta(\rho)}+
\frac{\mmu(\rho)\cdot \N}{\theta(\rho)} \ ,
\end{equation}
 where  $S(\rho)$ is the entropy operator [the precise definition is
 given in
Section \ref{Entropy}, Eq.\ (\ref{Sdef}), and the functionals
$\theta(\rho)$ and $\mmu(\rho)$ are defined in Section
\ref{Example}, Eqs.\ (\ref{Msteepest})-(\ref{constraint2})]. We
call operator $M(\rho)$
 the {\it generalized nonequilibrium Massieu-function operator}, because
  at thermodynamic equilibrium its mean
value belongs to the family of entropic characteristic functions
 introduced by Massieu \cite{Massieu}, i.e.,
\begin{equation}
\label{MassieuEq} \langle M\rangle_{\rm e}= \langle S\rangle_{\rm
e}-\frac{\langle H\rangle_{\rm e}}{T}+ \frac{\mmu\cdot
\langle\N\rangle_{\rm e}}{T} \ ,
\end{equation}
where $\langle S\rangle_{\rm e}$, $\langle H\rangle_{\rm e}$,
$\langle\N\rangle_{\rm e}$, $T$ and $\mmu$ are the equilibrium
entropy, energy, amounts of constituents, temperature and chemical
potentials, respectively.

The specific physical interpretations of the uncertainty relations
that we derive from dynamical law (\ref{rhodotHM}) depend on the
theoretical or modeling context in which it is assumed. In this
article we limit the discussion to generalities in Sections
\ref{General}, \ref{Entropy}, \ref{Shortest}, \ref{Occupation} and
to illustrative considerations and numerical results valid within
the simplest framework of steepest-entropy-ascent conservative
dynamics in Sections \ref{Example} and \ref{Numerical}.

\section{\label{General}General uncertainty relations}

We consider the space $\Ell (\Hil)$ of linear operators on $\Hil$
equipped with  the real scalar product
\begin{equation}
\label{Scalar} (F|G) = \Tr (F^\dagger G + G^\dagger F)/2 =(G|F) \
,
\end{equation}
and the real antisymmetric bilinear form
\begin{equation}
\label{Commutator} (F\backslash G) = i\,\Tr (F^\dagger G -
G^\dagger F)/2 =-(G\backslash F)= (F|iG)\ ,
\end{equation}
so that for any (time-independent) hermitian $F$ in $\Ell (\Hil)$
the corresponding mean-value state functional can be written as
$\mean{F}= \Tr (\rho F)=\Tr (\sq F\sq)=(\sq|\sq F)$, and can
therefore be viewed as a functional of $\sq$, the square-root
density operator, obtained from the spectral expansion of $\rho$
by substituting its eigenvalues with their positive square roots.
When $\rho$ evolves according to Eq. (\ref{rhodot}), the rate of
change of $\Tr (\rho F)$ can be written as
\begin{equation}
 \label{rateF}  {\rm d} \Tr (\rho F)/{\rm d}t = \Tr (
F\,{\rm d} \rho /{\rm d}t)=2\left.\left(\sq F\right|\sq
E(\rho)\right) \ .
\end{equation}

In particular, for the evolution equation (\ref{rhodot}) to be
well defined, the functional $\Tr (\rho I)$ where $I$ is the
identity on $\Hil$ must remain equal to unity at all times;
therefore, $ {\rm d} \Tr (\rho I)/{\rm d}t =2\left.\left(\sq
I\right|\sq E(\rho)\right)=0 $,  in view of Eq.\ (\ref{rhodotHM})
implies the condition
\begin{equation}
 \label{rateM}\left.\left(\sq \right|\sq \Delta M(\rho)\right) =0\ .
\end{equation}

For $F$ and $G$ hermitian in $\Ell (\Hil)$,  we introduce the
following shorthand notation
\begin{eqnarray}
\label{Delta}\Delta F&=&F- \Tr(\rho F)I \ ,\\
\label{Cov}\scov{F}{G}&=&\cov{F}{G}= (\sq \Delta F|\sq \Delta G)
\nonumber\\ &=&\frac{1}{2} \Tr(\rho\{\Delta F,\Delta
G\})=\scov{G}{F}\ ,\\ \Delta_F&=&\sqcov{F}{F}=\sqrt{\cov{F}{F}}\
,\\
  \label{Covi}\ceta{F}{G}&=&\com{F}{G}= (\sq
\Delta F\backslash \sq \Delta G) \nonumber\\&=& \frac{1}{2i}
\Tr(\rho[ F, G])=\cseta{F}{G}=-\ceta{G}{F}\ ,
\end{eqnarray}
 For
example, we may write the rate of change of the mean value of a
time-independent observable $F$ as
\begin{equation}
 \label{rateF2} \frac{{\rm d}\Tr (\rho F)}{{\rm d}t} =
 \frac{\com{F}{H}}{\hbar/2}
 + \frac{\cov{F}{M}}{\Boltz\tau}=\frac{\ceta{F}{H}}{\hbar/2}
 + \frac{\scov{F}{M}}{\Boltz\tau} \ ,
\end{equation}
from which we see that not all operators $F$ that commute with $H$
correspond to constants of the motion, but only those for which
$\cov{F}{M}=0$, i.e., such that $\sq\Delta F$ is orthogonal to
both $i\sq\Delta H$ and $\sq\Delta M$, in the sense of scalar
product (\ref{Scalar}). For an isolated system, conservation of
the mean energy functional $\Tr(\rho H)$ requires an operator
function $\Delta M(\rho)$ that maintains $\sq\Delta M$ always
orthogonal to $\sq\Delta H$, so that $\cov{H}{M}=0$  for every
$\rho$.

From Schwarz inequality, we readily verify \cite{proof1}
 the following generalized
Schr\"odinger uncertainty relation \cite{SchroedingerInequality}
\begin{equation}
 \label{Sinequality} \cov{F}{F}\cov{G}{G}\ge \cov{F}{G}^2+  \com{F}{G}^2\
 ,
\end{equation}
usually written in the form
$\sqrt{\scov{F}{F}\scov{G}{G}-\sccov{F}{G}}\ge |\ceta{F}{G}|$.
Relation (\ref{Sinequality}) obviously entails the less precise
and less symmetric Heisenberg-Robertson uncertainty relation
\begin{equation}
 \label{Rinequality} \cov{F}{F}\cov{G}{G}\ge  \com{F}{G}^2\ ,
\end{equation}
usually written in the form $\Delta_F\Delta_G\ge |\ceta{F}{G}|$.

For further compactness, we introduce the notation
\begin{eqnarray}
\label{rFG} r_{FG}&=&\scov{F}{G}\big/\sqrt{\scov{F}{F}\scov{G}{G}}
\ ,\nonumber\\ \label{cFG}
c_{FG}&=&\ceta{F}{G}\big/\sqrt{\scov{F}{F}\scov{G}{G}} \
,\end{eqnarray} where clearly, $r_{FG}$ represents the cosine of
the angle between the `vectors' $\sq \Delta F$ and $\sq \Delta G$
in $\Ell(\Hil)$, and $ r^2_{FG}\le 1$. Inequality
(\ref{Sinequality}) may thus be rewritten as
 \begin{equation}
 r^2_{FG}+c^2_{FG}\le 1 \end{equation} and
clearly implies \begin{equation}c^2_{FG}\le
\frac{1}{1+(r^2_{FG}/c^2_{FG})}\le 1-r^2_{FG}\le 1\
.\end{equation}

Next, for any hermitian $F$ we define the
 characteristic time of change  of the corresponding property defined by the  mean
value of the linear functional $\mean{F}= \Tr(\rho F)$  as follows
\begin{equation}
\tau_{F}(\rho)=\Delta_F\big/|{\rm d}\mean{F}/{\rm d}t| \
.\label{tauF}
\end{equation}
As is well known \cite{reviewsold,Messiah}, $\tau_{F}$ represents
the time required for the statistical distribution of measurements
of observable $F$ to be appreciably modified, i.e., for the mean
value $\mean{F}$ to change by an amount equal to the width
$\Delta_F$ of the distribution.

Now, defining the nonnegative, dimensionless functional
\begin{equation}\label{ataudef}
a_\tau=\hbar\Delta_M\Big/2\Boltz\tau\Delta_H \ ,
 \end{equation}
 we  rewrite (\ref{rateF2}) in the form
\begin{equation}
 \label{rateF3} {\rm d}\mean{F}/{\rm d}t =
 2\Delta_F\Delta_H\,(\cFH+a_\tau\,\rFM)/\hbar
 \end{equation}
and, substituting  into (\ref{tauF}), we obtain the general exact
uncertainty relation
\begin{equation}\label{exactTE}
\frac{\hbar/2}{\tau_{F}\Delta_H}= |\cFH+a_\tau\,\rFM| \ .
 \end{equation}
For non-dissipative dynamics, $a_\tau=0$, Eq.\ (\ref{exactTE})
yields the time-energy uncertainty relations
 \begin{equation}\label{nondissTE} \frac{\hbar^2/4}{\tau^2_{F}\scov{H}{H}}
 =\ccFH\le
\frac{1}{1+(\rrFH/\ccFH)}\le 1-\rrFH\le 1\ ,\end{equation}
 which entail but are more
precise than the usual time-energy uncertainty relation, in the
same sense as Schr\"odinger's relation (\ref{Sinequality}) entails
but is more precise than Heisenberg's relation
(\ref{Rinequality}). According to ({\ref{tauF}), the last
inequality in (\ref{nondissTE}) implies that property $\mean{F}$
cannot change at rates faster than $2\Delta_F\Delta_H/\hbar$.

For dissipative dynamics let us first consider an observable $A$
that commutes with $H$, so that $\com{A}{H}= 0$ while
$\cov{A}{H}\ne 0$; in other words, an observable conserved by the
Hamiltonian term in the dynamical law (\ref{rhodotHM}), but  not
conserved by the dissipative term. Then Eq.\ (\ref{exactTE})
yields the equivalent time-energy uncertainty relations
\begin{equation}\label{dissTEA}
\frac{\hbar/2}{\tau_{A}\Delta_H}= a_\tau\,|\rAM|\le a_\tau \ ,
 \end{equation}
\begin{equation}\label{dissTEA2}
\frac{\Boltz\tau(\rho)}{\tau_{A}\Delta_M}= |\rAM|\le 1 \ .
 \end{equation}

We note that while $\rrAM\le 1$, the value of $a_\tau$ depends on
how $\tau(\rho)$ is defined and, a priori, could well be larger
than unity, in which case there could be some observables $A$ for
which $\tau_{A}\Delta_H\le \hbar/2$. If instead we impose that the
functional $\tau(\rho)$ be defined in such a way that $a_\tau\le
1$, i.e.,
\begin{equation}\label{taubound}
\tau(\rho)\ge \hbar\Delta_M\Big/2\Boltz\Delta_H \ ,
 \end{equation}
 than we obtain that  even in dissipative
dynamics the usual time-energy uncertainty relations are never
violated by observables $A$ commuting with $H$.

However, in general, if the dynamics is dissipative
($\tau\ne\infty$) there are density operators for which
$|\cFH+a_\tau\,\rFM | >1$ so that $\tau_{F}\Delta_H$ takes a value
less than $\hbar/2$ and thus the usual time-energy uncertainty
relation is violated. The sharpest general time-energy uncertainty
relation always satisfied when both Hamiltonian and dissipative
dynamics are active is (proof in Section \ref{Shortest})
\begin{equation}\label{genunc1}
\frac{\hbar^2/4}{\tau^2_{F}\scov{H}{H}} \le 1+a_\tau^2+2a_\tau\cMH
\ ,
\end{equation}
which may also take the equivalent form
\begin{equation}\label{genunc2}
\frac{\tau^2_{F}\scov{H}{H}}{\hbar^2/4}+
\frac{\tau^2_{F}\scov{M}{M}}{\Boltz^2\tau^2(\rho)}+
\frac{\tau^2_{F}\Delta_M\Delta_H\cMH}{\Boltz\tau(\rho)\,\hbar/4}\ge
1 \ .
\end{equation}
 The upper bound in the rate of change of
 property $
\mean{F}$ becomes
\begin{equation}\label{bound}
\Delta_F\sqrt{\frac{\scov{H}{H}}{\hbar^2/4}+
\frac{\scov{M}{M}}{\Boltz^2\tau(\rho)}+
\frac{\Delta_M\Delta_H\cMH}{\Boltz\tau(\rho)\,\hbar/4}}\ .
\end{equation}
As anticipated,  because the dissipative term in Eq.\
(\ref{rhodotHM}) implies an additional dynamical mechanism, this
bound (\ref{bound}), valid for the particular nonunitary dynamics
we are considering, is higher than the standard bound valid in
unitary hamiltonian dynamics, given by $2\Delta_F\Delta_H/\hbar$.
For observables commuting with $H$, however, ({\ref{dissTEA2})
provides the sharper general bound $\Delta_F\Delta_M/\Boltz\tau$,
solely due to dissipative dynamics,  which is lower  than
({\ref{bound}).

Because in general $|\cMH|<1$, ({\ref{genunc2}) obviously implies
the less precise relation
\begin{equation}\label{genunc8}
\frac{\hbar^2/4}{\tau^2_{F}\scov{H}{H}} \le (1+a_\tau)^2 \ .
\end{equation}
However, as for the dynamics we discuss in Section \ref{Example},
if the Massieu operator function $\Delta M(\rho)$ is a linear
combination (with coefficients that may depend nonlinearly on
$\rho$) of operators that commute with either  $\rho$ or  $H$,
then it is easy to show that $\cMH=0$. Therefore, in such
important case, ({\ref{genunc2}) becomes
\begin{equation}\label{genunc9}
\frac{\hbar^2/4}{\tau^2_{F}\scov{H}{H}} \le 1+a_\tau^2 \ ,
\end{equation}
clearly sharper than ({\ref{genunc8}).  If in addition
$\tau(\rho)$ satisfies (\ref{taubound}), then (\ref{genunc9})
implies $\tau_{F}\Delta_H \ge \hbar/2\sqrt{2}$.

\section{\label{Entropy}Rate of entropy change characteristic time}

We now consider the entropy  functional $\mean{S}= \Tr(\rho S) =
-\Boltz\Tr (\rho\ln\rho) =-\Boltz \left(\sq \left|\sq
\ln(\sq)^2\right.\right)$ and its rate of change, which using
Eqs.\ (\ref{rhodotHM}) and (\ref{rateM}) may be written as
\begin{eqnarray} \label{rateS} {\rm d}\Tr(\rho S)/{\rm d}t &=&
2\left(\left.\sq S \right|\sq
E(\rho)\right)=\cov{S}{M}\big/\Boltz\tau\nonumber\\ &=&
\Delta_S\Delta_M\, \rSM\big/\Boltz\tau
\end{eqnarray}
where, for convenience, we define the entropy operator
\begin{equation}
\label{Sdef} S=-\Boltz P_{\rho>0}\ln\rho \ ,
\end{equation}
where $P_{\rho>0}$ is the projection operator onto the range of
$\rho$ \cite{operatorB}. Interestingly, the rate of entropy
change, being proportional to the correlation coefficient between
entropy measurements and $M$ measurements, under the assumptions
made so far, may be positive or negative, depending on how $M$ is
defined, i.e., depending on the specifics of the physical model in
which Eq.\ (\ref{rhodotHM}) is adopted.

The characteristic time of change of the entropy functional,
defined as
\begin{equation}
\tau_{S}(\rho)=\Delta_S\big/|{\rm d}\mean{S}/{\rm d}t| \
,\label{tauS}
\end{equation}
gives rise to the following equivalent exact time-energy
uncertainty relations
\begin{equation}
\frac{\hbar/2}{\tau_{S}\Delta_H}=a_\tau\,|\rSM | \le a_\tau \
,\label{TES}
\end{equation}
\begin{equation}\label{genunc3}
\frac{\Boltz\tau(\rho)}{\tau_{S}\Delta_M}=|\rSM | \le 1\ ,
\end{equation} where $r_{\!S\!M}$  is defined as in
(\ref{rFG}) using operators $\Delta M(\rho)$ and $\Delta
S=S-\mean{S}$. The physical interpretation of (\ref{genunc3}) is
that the entropy cannot change in time at a rate faster than
$\Delta_S\Delta_M/\Boltz\tau$, as immediately obvious also from
(\ref{rateS}).

We notice from (\ref{TES}) that if the dissipation time functional
$\tau(\rho)$ satisfies condition (\ref{taubound}) then $a_\tau\le
1$ and, therefore, the entropy change characteristic time $\tau_S$
satisfies the usual uncertainty relation $\tau_{S}\Delta_H\ge
\hbar/2$ and the rate of entropy change cannot exceed
$2\Delta_S\Delta_H/\hbar$.

We conclude this Section by noting that, in general, the equality
in (\ref{TES}) may be used to rewrite
 Relation (\ref{genunc1}) in the form
\begin{equation}\label{genunc4}
\frac{a_\tau}{1+a_\tau} |\rSM|\tau_S\le
\tau_F\frac{\sqrt{1+a^2_\tau+2 a_\tau\cMH}}{1+a_\tau}\le \tau_F\ ,
\end{equation}
where the last inequality follows from $|\cMH|\le 1$. This
relation shows, on one hand, that the entropy change
characteristic time $\tau_S$ is not necessarily the shortest among
the characteristic times $\tau_F$ associated with observables of
the type $\mean{F}=\Tr(\rho F)$ according to the Mandelstam-Tamm
definition (\ref{tauF}). On the other hand, it also shows that the
left-hand side defines a characteristic-time functional
\begin{equation}\label{tauHD}
\tau_{U\! D}=\frac{a_\tau}{1+a_\tau} |\rSM|\tau_S\le \tau_F\ ,
\end{equation}
which constitutes a general lower bound for all $\tau_F$'s, and
may therefore be considered the shortest characteristic time of
 simultaneous unitary/dissipative dynamics as described by
Equation (\ref{rhodotHM}). This observation prompts the discussion
in the next Section.

\section{\label{Shortest}Shortest characteristic times for
purely-unitary and purely-dissipative dynamics}

The Mandelstam-Tamm definition (\ref{tauF}) of characteristic
times has been criticized for various reasons (see for example
Refs.\ \citen{Eberly,Leubner,Bhattacharyya}) mainly related to the
fact that depending on which observable $F$ is investigated, as
seen by inspecting (\ref{nondissTE}), the bound $\tau_F\ge
\hbar/2\Delta_H$ may be very poor whenever $\ccFH$ is much smaller
than 1.

Therefore, different attempts have been made to define
characteristic times that (1) refer to the quantum system as a
whole rather than to some particular observable, and (2) bound all
the particular $\tau_F$'s from below. Notable examples are the
characteristic times $\tau_{ES}$ and $\tau_{LK}$, respectively
defined by Eberly and Singh \cite{Eberly} and  Leubner and Kiener
\cite{Leubner} as follows

The following definitions are based on the observation that
$\Delta_F$ may be interpreted as the norm of $\sq \Delta F$
(viewed as a vector in $\Ell (\Hil)$) in the sense that it equals
$\sqrt{(\sq \Delta F |\sq\Delta F)}$, therefore, we may use it to
define the (generally non hermitian) unit norm vector in $\Ell
(\Hil)$
\begin{equation}
\label{normalized}  \tilde F_\rho =\sq \Delta F/\Delta_F \ .
\end{equation}
As a result, Eq. (\ref{rateF2}) may be rewritten in the form
\begin{equation}
 \label{rateFtilde} \frac{1}{\Delta_F} \frac{ {\rm d} \mean{F}}{{\rm d}t}
 =\frac{\Delta_H}{\hbar/2}\,(\tilde F_\rho|i\tilde H_\rho) +
  \frac{\Delta_M}{\Boltz \tau}\,(\tilde F_\rho|\tilde M_\rho)= (\tilde F_\rho|C)\ ,
\end{equation}
where for shorthand we define the  operator
\begin{equation}
\label{defC}  C = i\frac{\Delta_H \tilde
H_\rho}{\hbar/2}+\frac{\Delta_M \tilde M_\rho}{\Boltz\tau}=2\sq
E(\rho) \ ,
\end{equation}
directly related [see Eq. (\ref{rateF})] with the evolution
operator function $E(\rho)$ defined in the Introduction, which
determines the rates of change of all linear
 functionals of the state operator $\rho$, i.e., all observables of
 the linear type $\Tr(\rho F)$, by  its
   projection onto the respective directions $\tilde F_\rho$.

Each characteristic time $\tau_F$ can now be written as
\begin{equation}
\label{tauF2}  \tau_F=\Delta_F/|{\rm d} \mean{F}/dt|=1/|(\tilde
F_\rho|C)| \   .
\end{equation}
Because $\tilde F_\rho$ is unit norm, $|(\tilde F_\rho|C)|$ is
bounded by the value attained for an operator $\tilde F_\rho$ that
has the same `direction' in $\Ell (\Hil)$ as operator $C$, i.e.,
for
\begin{equation}
\label{frho} \tilde F_\rho = \pm C/\sqrt{(C|C)} \   ,
\end{equation}
in which case $|(\tilde
F_\rho|C)|=\sqrt{(C|C)}=\sqrt{\Tr(C^\dagger C)}$. Thus we conclude
that, for any, $F$,
\begin{equation}
\label{ineqCF}1/\sqrt{(C|C)}\le \tau_F \   ,
\end{equation}
and, therefore, we introduce the shortest characteristic time for
the combined unitary-dissipative dynamics described by Eq.
(\ref{rhodotHM}),
\begin{equation}
\label{tauUD}\tau_{U\! D}=1/\sqrt{(C|C)} \   ,
\end{equation}
which bounds from below all $\tau_F$'s. From (\ref{defC}) and
(\ref{ineqCF}), and the identities $(i\tilde H_\rho |i\tilde
H_\rho)=(\tilde M_\rho |\tilde M_\rho)=1$ and $(i\tilde H_\rho
|\tilde M_\rho)=(\tilde M_\rho |i\tilde H_\rho)=\cMH$ we obtain
\begin{eqnarray}
\label{ineqUD} \frac{1}{\tau_F^2}\le\frac{1}{\tau_{U\!
D}^2}&=&(C|C)= \frac{\scov{H}{H}}{\hbar^2/4}+
\frac{\scov{M}{M}}{\Boltz^2\tau^2(\rho)}+
\frac{\Delta_M\Delta_H\cMH}{\Boltz\tau(\rho)\,\hbar/4}\nonumber\\
&=&\frac{\scov{H}{H}}{\hbar^2/4}(1+a_\tau^2+2\,a_\tau\cMH) \ ,
\end{eqnarray}
which proves Relations (\ref{genunc1}) and (\ref{genunc2}).

For nondissipative (purely Hamiltonian, unitary) dynamics the same
reasoning (or substitution of $\tau=\infty$, $a_\tau=0$ in the
above relations) leads to the definition of the shortest
characteristic time
\begin{equation}
\label{deftauU}\tau_U=\hbar/2\Delta_H \   ,
\end{equation}
with which the usual time-energy relation reduces to
\begin{equation} \label{genunc11}\tau_F\ge \tau_U \   .
\end{equation}
Its physical meaning is that when the energy dispersion (or
uncertainty or spread) $\Delta_H$ is small, $\tau_U$ is large and
$\tau_F$ must be larger for all observables $F$, therefore, the
mean values of all properties change slowly \cite{Pfeifer}, i.e.,
the state $\rho$ has a long lifetime. Conversely, states with a
small energy spread cannot change rapidly with time. States that
change rapidly due to unitary dynamics, necessarily have a large
energy spread.

Another interesting extreme case obtains from Eq. (\ref{rhodotHM})
when $\Delta M(\rho)$ is such that the condition $[\rho,H]=0$
implies $[\Delta M(\rho),H]=0$ for any $\rho$, as for the
steepest-entropy-ascent dynamics discussed in Sections
\ref{Example} and \ref{Numerical}. In this case, it is easy to see
that if the state operator $\rho$ commutes with $H$ at one instant
of time then it commutes with $H$ at all times and, therefore, the
entire time evolution is purely dissipative. Then, the reasoning
above leads to the definition of the shortest characteristic time
\begin{equation}
\label{deftauD}\tau_D=\Boltz\tau/\Delta_M \   .
\end{equation}
It s noteworthy that $\tau_D$ can be viewed as the characteristic
time associated not with the (generally nonlinear) Massieu
functional $\mean{M}=\Tr(\rho M(\rho))$ but with the linear
functional $\mean{A}=\Tr(\rho A)$ corresponding to the
time-independent operator $A$ which at time $t$ happens to
coincide with $M(\rho(t))$.

For purely dissipative dynamics, the bound
$\tau_F\ge\tau_D=\Boltz\tau/\Delta_M$ implies that when
$\Delta_M/\Boltz\tau$, i.e., the ratio between the spread in our
generalized nonequilibrium Massieu function and the dissipation
time functional, is small, then $\tau_D$ is large and $\tau_F$
must be larger for all observables $F$, therefore, the state
$\rho$ has a long lifetime. Conversely, if some observable changes
rapidly, $\tau_F$ is small and since $\tau_D$ must be smaller, we
conclude that the spread $\Delta_M$ [more precisely, the ratio
$\Delta_M(\rho)/\Boltz\tau(\rho)$] must be large.

In terms of $\tau_U$ and $\tau_D$ we can rewrite (\ref{ataudef}),
(\ref{genunc3}) and (\ref{ineqUD}) as
\begin{equation}
a_\tau=\tau_U/\tau_D \   ,
\end{equation}
\begin{equation}\label{tauSD}
\frac{1}{\tau_S}=\frac{|\rSM|}{\tau_D}\le \frac{1}{\tau_D} \   ,
\end{equation}
\begin{eqnarray}
\label{ineqUD2}
\frac{1}{\tau_F^2}&=&\left(\frac{\cFH}{\tau_U}+\frac{\rFM}{\tau_D}\right)^2
\nonumber\\ &\le&\frac{1}{\tau_{U\!
D}^2}=\frac{1}{\tau_U^2}+\frac{1}{\tau_D^2}
+\frac{2\,\cMH}{\tau_U\tau_D} \nonumber\\
&\le&\left(\frac{1}{\tau_U}+\frac{1}{\tau_D}\right)^2  .
\end{eqnarray}
Eq. (\ref{tauSD}) implies that the entropy cannot change rapidly
with time  if the ratio $\Delta_M(\rho)/\Boltz\tau(\rho)$ is not
large. The first equality in (\ref{ineqUD2}) follows from $(\tilde
F_\rho |i\tilde H_\rho)=\cFH$ and $(\tilde F_\rho |\tilde
M_\rho)=\rFM$, which also imply that  Eq. (\ref{rateFtilde}) may
take the form
\begin{equation}
 \label{rateFtilde2}   \frac{{\rm d} \mean{F}}{{\rm d}t}
  =\Delta_F\left(\frac{\cFH}{\tau_U} +
  \frac{\rFM}{\tau_D}\right)
 \ ,
\end{equation}
and operator $C$ defined in (\ref{defC}) takes also the forms
\begin{equation}
C= i\frac{\tilde H_\rho}{\tau_U} +
 \frac{\tilde M_\rho}{\tau_D}= i\frac{\sq\Delta H}{\Delta_H\tau_U}+\frac{\sq\Delta
 M}{\Delta_M\tau_D} \ ,
 \end{equation} and its norm is
 $\sqrt{1/\tau_U^2+1/\tau_D^2+2\cMH/\tau_U\tau_D}$.

 Similarly, the rate of entropy change (\ref{rateS}) takes the
 form
\begin{equation} \label{rateS2} \frac{{\rm d}\mean{S}}{{\rm d}t} =
\frac{\Delta_S}{\tau_D}\,(\tilde S_\rho\,|\,\tilde
M_\rho)=\frac{\Delta_S\, \rSM}{\tau_D}
\end{equation}
which, because $|\rSM |\le 1$, implies the bounds [equivalent to
(\ref{genunc3}) and (\ref{tauSD})],
\begin{equation} \label{rateSbound} -\frac{\Delta_S}{\tau_D}\le
\frac{{\rm d}\mean{S}}{{\rm d}t} \le \frac{\Delta_S}{\tau_D} \ .
\end{equation}

\section{\label{Occupation}Occupation probabilities}

An important class of observables for a quantum system are those
associated with the projection operators. For example,
 for pure states evolving unitarily \cite{Pfeifer}, the mean value
 $\mean{P}=\Tr(\rho(t)
P)$ where $P=|\phi_0\rangle\langle\phi_0|=\rho(0)$ represents the
survival probability of the initial state, and is related to
several notions of lifetimes \cite{Pfeifer}.

We do not restrict our attention to pure states, and we discuss
 first  results that hold for any projector $P$ associated with a
 yes/no type of measurement. Let $P=P^\dagger=P^2$ be an
 orthogonal projector onto the $g$-dimensional subspace $P\Hil$ of
 $\Hil$. Clearly, $g=\Tr(P)$, the variance
 $\cov{P}{P}=p\,(1-p)$ where $p=\mean{P}=\Tr(\rho
P)$ denotes the mean value and represents the probability in state
$\rho$ of obtaining a `yes'  result upon measuring the associated
observable, and the characteristic time of the rate of change of
this occupation probability is defined according to (\ref{tauF})
by
\begin{eqnarray}
\label{tauP}\frac{1}{\tau_{P}}&=&\frac{|{\rm d}p/{\rm
d}t|}{\sqrt{p\,(1-p)}}=2\left|\frac{\rm d}{{\rm
d}t}\arccos(\sqrt{p})\right|\nonumber\\&=& 2\left|\frac{\rm
d}{{\rm d}t}\arcsin(\sqrt{p})\right|\le\frac{1}{\tau_{U\! D}}\ ,
\end{eqnarray}
where the inequality follows from (\ref{ineqUD}). Therefore,
\begin{equation}
\label{cospn}-\frac{1}{2\tau_{U\! D}}\le\frac{\rm d}{{\rm
d}t}\arccos(\sqrt{p})\le \frac{1}{2\tau_{U\! D}} \ ,
\end{equation}
or, over any finite time interval of any time history $p(t)$,
\begin{equation}
\label{cosfinite}\left|\arccos(\sqrt{p(t_2)})-\arccos(\sqrt{p(t_1)})\right|\le
\left|\int_{t_1}^{t_2}{\frac{{\rm d}t'}{2\tau_{U\! D}(t')}}\right|
\ .
\end{equation}
This result generalizes the results on lifetimes  obtained in
\cite{Bhattacharyya} where the focus is restricted to full quantum
decay [$p(\infty)\approx 0$] of an initially fully populated state
[$p(0)\approx 1$] and $\tau_U$ (here $\tau_{U\! D}$) is assumed
constant during the time interval. It is also directly related to
some of the results in \cite{Pfeifer}, where a number of
additional inequalities and bounds on lifetimes are obtained for
unitary dynamics, and may be straightforwardly generalized to the
class of simultaneous unitary/dissipative dynamics  described by
our Eq. (\ref{rhodotHM}).

Because $p\,(1-p)$ attains its maximum value when $p=1/2$, we also
have the inequality
\begin{equation}
\label{dpndt}\left|\frac{{\rm d}p}{{\rm d}t}\right|\le
\frac{1}{2\tau_{U\! D}} \ .
\end{equation}
which, analogously to what noted in  \cite{Bhattacharyya}, implies
that no full decay nor full population can occur within a time
$2\tau_{U\! D}$, so that this time may be interpreted as a limit
to the degree of instability of a quantum state.

Next, we focus on the projectors onto the eigenspaces of the
Hamiltonian operator $H$, assumed time-independent. Let us write
its spectral expansion as $H=\sum_n e_n\Pen$ where $e_n$ is the
$n$-th eigenvalue and $\Pen$ the projector onto the corresponding
eigenspace. Clearly, $H\Pen =e_n\Pen$, $\Pen\Pem=\delta_{nm}\Pen$,
 $g_n=\Tr(\Pen )$ is the degeneracy of eigenvalue
$e_n$, $p_n=\mean{\Pen }=\Tr(\rho \Pen )$ the occupation
probability of energy level $e_n$, $\cov{\Pen }{\Pem
}=p_n\,(\delta_{nm}-p_m)$ the covariance of pairs of occupations,
and $\cov{\Pen }{\Pen }=p_n\,(1-p_n)$ the variance or fluctuation
of the $n$-th occupation. Because $[\Pen ,H]=0$, $\cPenH=0$, by
(\ref{rateFtilde2}) we have
\begin{equation} \label{ratePen}   \frac{{\rm d} p_n}{{\rm d}t}
  =\Delta_{\Pen}
  \frac{\rPenM}{\tau_D}
 \ ,\end{equation}
 and the corresponding characteristic time is
 \begin{equation} \label{teuPen}   \frac{1}{\tau_\Pen}
  =
  \frac{|\rPenM |}{\tau_D} \le \frac{1}{\tau_D}
 \ .\end{equation}

Energy level occupation probabilities $p_n$ are used in Section
\ref{Numerical} for numerical illustration/validation of
inequalities (\ref{teuPen}) within the steepest entropy ascent
dynamical model outlined in the next Section.

\section{\label{Example}Example. Steepest-entropy-ascent,
conservative dissipative dynamics}

So far we have not assumed an explicit form of $\Delta M(\rho)$
except for condition (\ref{rateM})  that maintains $\rho$ unit
trace. In this section, we illustrate the above results  by
further assuming steepest-entropy-ascent, conservative dissipative
dynamics as obtained by assuming for our generalized
nonequilibrium Massieu operator function the expression
\begin{equation}
\label{Msteepest} \Delta M(\rho)=\Delta S- \Delta
H'(\rho)/\theta(\rho) \ ,
\end{equation}
where $S$ is the entropy operator defined in (\ref{Sdef}),
\begin{equation}
\label{defH'} \Delta H'(\rho) = \Delta H - \mmu(\rho)\cdot
\Delta\N\ ,
\end{equation}
 $H$ is the
Hamiltonian, $\N=\{N_1,\dots,N_r\}$  a (possibly empty) set of
operators  commuting with $H$,   that we call non-Hamiltonian
generators of the motion (for example, the number-of-particles
operators or a subset of them, or the momentum component operators
for a free particle),  such that operators $\sq\Delta H$ and $\sq
\Delta \N$ are linearly independent, $\theta(\rho)$ and
$\mmu(\rho)=\{\mu_1(\rho),\dots,\mu_r(\rho)\}$ a set of real
functionals defined for each $\rho$ by the solution of the
following system of linear equations
\begin{eqnarray}
\label{constraint1}
\cov{S}{H}\,\theta+\sum_{i=1}^r\cov{N_i}{H}\,\mu_i&=& \cov{H}{H}\
,\\ \label{constraint2}
\cov{S}{N_j}\,\theta+\sum_{i=1}^r\cov{N_i}{N_j}\,\mu_i&=&
\cov{H}{N_j} \ ,
\end{eqnarray}
which warrant the conditions that $\cov{H}{M}=0$ and
$\cov{N_j}{M}=0$, and hence that the mean values $\Tr(\rho H)$ and
$\Tr(\rho \N)$ are maintained time invariant by the dissipative
term of the equation of the motion.

Operators $\sq\Delta H'$ and $\sq\Delta M$ are always orthogonal,
in the sense that $\cov{M}{H'}=0$ for every $\rho$. It follows
that, in general, $\cov{S}{H'}=\cov{H'}{H'}/\theta$,
\begin{equation}
\label{covSM} \cov{S}{M}= \cov{M}{M}=
\cov{S}{S}-\frac{\cov{H'}{H'}}{\theta^2(\rho)}\ge 0 \ ,
\end{equation}
and hence the rate of entropy generation (\ref{rateS}) is always
strictly positive except for $\cov{M}{M}=0$ (which occurs iff
$\sq\Delta M=0$), i.e., for$ \sqnd\Delta S_{\rm nd}=(\sqnd\Delta H
- \mmu_{\rm nd}\cdot\sqnd\Delta\N)/\theta_{\rm nd}$, for some real
scalars $\theta_{\rm nd}$ and $\mmu_{\rm nd}$, that is, for
density operators (that we call non-dissipative) of the form
\begin{equation}
\label{rhonondiss} \rho_{\rm nd}=\frac{B\exp[-( H - \mmu_{\rm
nd}\cdot\N)/\Boltz\theta_{\rm nd}]B}{\Tr B\exp[-( H - \mmu_{\rm
nd}\cdot\N)/\Boltz\theta_{\rm nd}]} \ ,
\end{equation}
where $B$ is any projection operator on $\Hil$ ($B^2=B$).

The nonlinear functional
\begin{equation}\label{theta}
\theta(\rho)=\frac{\scov{H'}{H'}}{\scov{S}{H'}}
=\frac{\Delta_{H'}}{\Delta_S\,\rSHprime}
\end{equation}
 may be interpreted in this
framework as a natural generalization to nonequilibrium of the
 temperature, at least insofar as for $t\rightarrow +\infty
$, while the state operator $\rho(t)$ approaches a non-dissipative
operator of form  (\ref{rhonondiss}), $\theta(t)$ approaches
smoothly the corresponding  thermodynamic equilibrium (or partial
equilibrium) temperature $\theta_{\rm nd}$.

Because here $H$ always commutes with $M$,  $\cMH=0$ and $(\tilde
M|i\tilde H)=0$, which means that $\sq \Delta M(\rho)$ is always
orthogonal to $i\sq \Delta H$. This reflects the fact that on the
entropy surface the direction of steepest entropy ascent is
orthogonal to the (constant entropy) orbits that characterize
purely
 Hamiltonian (unitary) motion (in which the entropy is maintained constant by
 keeping  invariant each eigenvalue of
 $\rho$).

Inequality (\ref{covSM}), which follows from $\rrSM\le 1$, implies
that
    $\scov{M}{M}\le \scov{S}{S}
  $ and $0\le\rSM=\Delta_M/\Delta_S\le 1$ or, equivalently,
\begin{equation}
\label{deftauG}\tau_K=\Boltz\tau/\Delta_S\le \tau_D \   ,
\end{equation}
where for convenience we  define the
   characteristic time $\tau_K$, which is simply related to the entropy spread, but
   cannot be attained by any rate of change, being shorter than
   $\tau_D$.
In addition, we have the identities
\begin{equation}
\label{rrSM}
\rrSM=\frac{\scov{M}{M}}{\scov{S}{S}}=\frac{\tau_K^2}{\tau_D^2}
=\frac{\tau_K}{\tau_S}
=1-\frac{\scov{H'}{H'}}{\theta^2\scov{S}{S}}=1-\rrSHprime \ ,
\end{equation}
and, from $\rrSHprime\le 1$, the bounds
\begin{equation}
\label{thetabound} |\theta| \ge \frac{\Delta_{H'}}{\Delta_S} \quad
{\rm or}\quad
-\frac{\Delta_S}{\Delta_{H'}}\le\frac{1}{\theta}\le\frac{\Delta_S}{\Delta_{H'}}
\ ,
\end{equation}
where the equality $ |\theta| =\Delta_{H'}/\Delta_S$ holds when
and only when the state is non-dissipative [Eq.
(\ref{rhonondiss})].  Additional bounds on our generalized
nonequilibrium temperature $\theta$ obtain by combining
(\ref{rrSM}) with the inequality $4\rrSM(1-\rrSM)\le 1$ (which
clearly holds because $ \rrSM\le 1$), to obtain
$4\rrSM\rrSHprime\le 1$ and, therefore,
\begin{equation}
\label{thetabound2}
\frac{2\Delta_M\Delta_{H'}}{|\theta|\scov{S}{S}}\le 1 \quad {\rm
or}\quad
-\frac{\scov{S}{S}}{2\Delta_M\Delta_{H'}}\le\frac{1}{\theta}\le\frac{\scov{S}{S}}{2\Delta_M\Delta_{H'}}
\ .
\end{equation}
At equilibrium,  $\Delta_M=0$ and (\ref{thetabound2}) implies no
actual bound on $\theta$, but in nonequilibrium states bounds
(\ref{thetabound2}) may be tighter than (\ref{thetabound}), as
illustrated by the numerical example in Section \ref{Numerical}.

   Notice that whereas in steepest entropy ascent dynamics
   $\tau_K$is always shorter than $\tau_D$ and
   obeys
   the identity \begin{equation}\label{identity}\tau_S\tau_K=\tau_D^2\ ,\end{equation}
 in general it is not necessarily shorter than $\tau_D$ and
obeys the identity
\begin{equation}\label{identitygen}\frac{\Delta_M}{\Delta_S}\frac{\tau_D^2}{\tau_S\tau_K}=|\rSM|\ ,\end{equation}

In summary, we conclude that
  within steepest-entropy-ascent, conservative dissipative quantum
 dynamics, the general uncertainty relations
 (\ref{genunc2}), (\ref{TES}) and (\ref{genunc3})
 that constitute the main results of this paper,
yield the
 time-energy/time-Massieu uncertainty relation
 \begin{equation}\label{genunc6M}
\left(\frac{\tau_{F}\Delta_H}{\hbar/2}\right)^2+
\left(\frac{\tau_{F}\Delta_M}{\Boltz\tau(\rho)}\right)^2
 \ge 1 \quad {\rm or}\quad
   \frac{\tau^2_{F}}{\tau^2_{U}}+\frac{\tau^2_{F}}{\tau^2_{D}} \ge 1 \ ,
\end{equation}
which implies the  interesting
 time-energy/time-entropy uncertainty relation
 \begin{equation}\label{genunc6}
\left(\frac{\tau_{F}\Delta_H}{\hbar/2}\right)^2+
\left(\frac{\tau_{F}\Delta_S}{\Boltz\tau(\rho)}\right)^2 \ge 1
\quad {\rm or}\quad
   \frac{\tau^2_{F}}{\tau^2_{U}}+\frac{\tau^2_{F}}{\tau^2_{K}} \ge 1 \ ,
\end{equation}
 and the  time-entropy  uncertainty relation
\begin{equation}\label{genunc5}
\frac{\tau_K}{\tau_{S}}=\frac{\Boltz\tau(\rho)}{\tau_{S}\Delta_S}=\rrSM
\le \rSM \le 1 \ ,
\end{equation}
which implies that the rate of entropy generation never exceeds
$\scov{S}{S}/\Boltz\tau$, i.e.,
\begin{equation}\label{genunc7}
\frac{{\rm d}\mean{S}}{{\rm d}t}=-\Boltz\frac{{\rm d}}{{\rm
d}t}\Tr (\rho \ln\rho) =\frac{\scov{M}{M}}{\Boltz\tau}
\le\frac{\Delta_S\Delta_M}{\Boltz\tau}
\le\frac{\scov{S}{S}}{\Boltz\tau}\ .
\end{equation}
If in addition the dynamics is purely dissipative, such as along a
trajectory $\rho(t)$ that commutes with $H$ for every $t$, then
(\ref{genunc6}) may be replaced by the time-entropy uncertainty
relation
 \begin{equation}\label{genunc12}
\frac{\tau_K}{\tau_{F}}=\frac{\Boltz\tau(\rho)}{\tau_{F}\Delta_S}
\le 1 \ .
\end{equation}

As shown in Refs.\ \citen{Beretta}, the dissipative dynamics
generated by Eq.\ (\ref{rhodotHM}) with $\Delta M(\rho)$ as just
defined and a time-independent Hamiltonian $H$: (i) maintains
$\rho(t)\ge \rho^2(t)$ at all times, both forward and backward in
time for any initial density operator $\rho(0)$ (see also
\cite{Gheorghiu}); (ii) maintains the cardinality of $\rho(t)$
invariant; (iii) entails that the entropy functional is an
$S$-function in the sense defined in \cite{Lyapunov} and therefore
that maximal entropy density operators obtained from
(\ref{rhonondiss}) with $B=I$ are the only
 equilibrium states of the dynamics that are stable with respect
 to perturbations that do not alter the mean values of the energy
 and the other time invariants (if any): this theorem of the
 dynamics  coincides with a well-known general statement of
 the second law of thermodynamics \cite{Book};
(iv)  entails Onsager reciprocity in the sense defined in
 \cite{Onsager}; (v) can be derived from a variational principle
 \cite{Gheorghiu}, equivalent to our steepest entropy ascent
 geometrical construction \cite{proofmaximal},
 by maximizing the entropy generation rate subject to the $\Tr(\rho)$,
  $\Tr(\rho H)$, and $\Tr(\rho\N)$ conservation  constraints and
  the additional constraint $(\sq E|\sq E)=c(\rho)$ \cite{proofmaximal}.

We finally note that assuming in Eq. (\ref{rhodotHM}), in addition
to $\Delta M(\rho)$ given by (\ref{Msteepest}), the nonlinear
relaxation time $\tau(\rho)$ given by (\ref{taubound}) with strict
equality, we obtain the most dissipative (maximal entropy
generation rate \cite{ArXiv1}) dynamics in which the entropic
characteristic time $\tau_S$ [Eq.\ (\ref{tauS})] is always
compatible with the time-energy uncertainty relation
$\tau_S\Delta_H\ge \hbar/2$ and the rate of entropy generation is
always given by $2\Delta_M\Delta_H/\hbar$.

The physical meaning of relations (\ref{genunc2}), (\ref{TES}),
(\ref{genunc3}), (\ref{genunc5}), (\ref{genunc6}) are worth
further investigations and experimental validation in specific
contexts in which the dissipative behavior is correctly modeled by
a dynamical law of form (\ref{rhodotHM}), possibly with $\Delta
M(\rho)$ of form (\ref{Msteepest}). One such context may be the
currently debated so-called ``fluctuation theorems" \cite{Crooks}
whereby fluctuations and, hence, uncertainties are measured on a
microscopic system (optically trapped colloidal particle
\cite{Wang}, electrical resistor \cite{Garnier}) driven at steady
state (off thermodynamic equilibrium) by means of a work
interaction, while a heat interaction (with a bath) removes the
entropy being generated by irreversibility. Another such context
may be that of pion-nucleus scattering, where available
experimental data have recently allowed partial validation
\cite{Ion} of ``entropic" uncertainty relations \cite{Deutsch}.
Yet another is within the model we propose in Ref.\ \citen{PRE}
for the description of  the irreversible time evolution of  a
perturbed, isolated, physical system during relaxation toward
thermodynamic equilibrium by spontaneous internal rearrangement of
the occupation probabilities. We pursue this example in the next
section.

\section{\label{Numerical}Numerical results for relaxation within
a dilute Boltzmann
gas of $N$-level particles}

To illustrate the time dependence of the uncertainty relations
derived in this paper, we consider an isolated, closed system
composed of noninteracting identical particles with
single-particle eigenstates with energies $e_i$ for $i=1$,
2,\dots, $N$,  where $N$ is assumed finite for simplicity and the
$e_i$'s are repeated in case of degeneracy, and we restrict our
attention to the class of dilute-Boltzmann-gas states in which the
particles are independently distributed among the $N$ (possibly
degenerate) one-particle energy eigenstates. This model is
introduced in Ref.\ \citen{PRE}, where we assume an equation of
 form (\ref{rhodotHM}) with $\Delta M(\rho)$ given by (\ref{Msteepest}) with the further simplification that
$\Delta H'(\rho) = \Delta H$ so that our generalized
nonequilibrium Massieu-function operator is
\begin{equation}
\label{MsteepestH} M(\rho)= S-  H/\theta(\rho) \ ,
\end{equation}
and, therefore,
\begin{equation}
\label{deltaMsteepestH} \Delta M(\rho)=\Delta S- \Delta
H/\theta(\rho) \ .
\end{equation}

 For simplicity and illustrative purposes, we focus on purely dissipative
 dynamics by considering a particular
trajectory $\rho(t)$ that commutes with $H$ at all times $t$,
assuming that $H$ is
 time independent and has a nondegenerate
spectrum. As a result, the energy-level occupation probabilities
$p_n$ coincide with the eigenvalues of $\rho$, and the dynamical
equation reduces to the simple form \cite{PRE}
\begin{equation}
\label{pdot} \frac{{\rm d}p_n}{{\rm
d}t}=-\frac{1}{\tau}\left[p_n\ln p_n +
p_n\frac{\mean{S}}{\Boltz}+p_n\frac{e_n-\mean{H}}{\Boltz\theta}\right]
\ ,
\end{equation}
where
\begin{eqnarray}
\label{pdotdefs} \mean{S}&=&-\Boltz\sum_n p_n\ln p_n \ ,\\
\mean{H}&=&\sum_n p_ne_n \ ,\\ \theta&=&\scov{H}{H}/\scov{H}{S} \
,\\ \scov{H}{H}&=&\sum_n p_ne^2_n-\mean{H}^2 \ ,\\
\scov{H}{S}&=&-\Boltz\sum_n p_ne_n\ln p_n -\mean{H}\mean{S} \ .
\end{eqnarray}

To obtain the plots in Figures 1-4, that illustrate the main
inequalities derived in this paper for a sample trajectory, we
consider  an initial state with cardinality equal to 4, with
nonzero occupation probabilities only for the four energy levels
$e_1=0$, $e_2=u/3$, $e_3=u/3$, and $e_4=u$, and with mean energy
$\mean{H}=2u/5$ ($u$ is arbitrary, with units of energy).
Moreover, as done in \cite{PRE}, we select an  initial state
$\rho(0)$ at time $t=0$ such that the resulting trajectory
$\rho(t)$ passes in the neighborhood of the partially canonical
nondissipative state $\rho^{\rm ft}_{\rm nd}$ that has  nonzero
occupation probabilities only for the three energy levels $e_1$,
$e_2$, and $e_4$, and mean energy $\mean{H}=2u/5$ ($p^{\rm
ft}_{{\rm nd}1}=0.3725$, $ p^{\rm ft}_{{\rm nd}2}=0.3412$, $p^{\rm
ft}_{{\rm nd}3}=0$, $p^{\rm ft}_{{\rm nd}4}=0.2863$, $\theta^{\rm
ft}_{\rm nd} =3.796\, u/\Boltz$ ). As shown in Figure 1, during
the first part of the trajectory, this nondissipative state
appears as an attractor, an approximate or `false target'
equilibrium state; when the trajectory gets close to this state,
the evolution slows down, the entropy generation drops almost to
zero and the value of $\theta$ gets very close ($3.767\,
u/\Boltz$) to that of $\theta^{\rm ft}_{\rm nd}$; however
eventually the small, but nonzero initial occupation of level
$e_3$ builds up and a new rapid rearrangement of the occupation
probabilities takes place, and finally drives the system toward
the maximal entropy state $\rho^{\rm pe}_{\rm nd}$ with energy
$\mean{H}=2u/5$ and all four active levels occupied, with
partially canonical distribution $p^{\rm pe}_{{\rm nd}1}=0.3474$,
$ p^{\rm pe}_{{\rm nd}2}=0.2722$, $p^{\rm pe}_{{\rm nd}3}=0.2133$,
$p^{\rm pe}_{{\rm nd}4}=0.1671$, and  characterized according to
(\ref{rhonondiss}) by the (partial equilibrium) temperature
$\theta^{\rm pe}_{\rm nd} =1.366\, u/\Boltz$.

The trajectory is computed by  integrating Eq.\ (\ref{pdot})
numerically, both forward and backward in time,
 starting from the chosen initial state $\rho(0)$, and assuming
 for Figures 1 and 2
 that the dissipation time $\tau$ is a constant, and for Figures 3
 and 4 that it is given by (\ref{taubound}) with strict equality
 ($a_\tau=1$, $\tau_D=\tau_U$), i.e, assuming
\begin{eqnarray}
\label{pdotdefs2}
\tau&=&\frac{\hbar/2}{\Boltz}\frac{\Delta_M}{\Delta_H}=
\frac{\hbar/2}{\Boltz}\sqrt{\frac{\scov{S}{S}}{\scov{H}{H}}-\frac{1}{\theta^2}}
\ ,\\ \scov{S}{S}&=&\Boltz^2\sum_n p_n(\ln p_n)^2-\mean{S}^2 \ .
\end{eqnarray}

The system of ordinary differential
 equations (\ref{pdot}) is highly nonlinear, especially when $\tau$
  is assumed according to
 (\ref{pdotdefs2}), nevertheless it is sufficiently well
 behaved to allow simple integration by means of a standard
 Runge-Kutta numerical scheme. Of course, we check that at all
 times $-\infty<t<\infty$ each $p_n$ remains nonnegative, $\sum_n
 p_n$ remains equal to unity, $\sum_n p_ne_n$ remains constant at
 the value $2u/5$ fixed by the selected initial state, and the
 rate of change of $\mean{S}$ is always nonnegative.

  In  each Figure, the top subfigure shows for ease of comparison the
  plots of the four nonzero occupation probabilities as functions of dimensionless
  time:  $t/\tau$, in Figures 1 and 2; $u\, t/\hbar$, in Figures 3
  and 4.
  The dots on the right represent the maximal
entropy distribution, $p_n(+\infty)=p_{\rm nd}^{\rm pe}$; the dots
at the left represent the lowest-entropy or `primordial'
distribution, $p_n(-\infty)=p_{\rm nd}^{\rm ls}$, which for the
particular trajectory selected here, corresponds to a
nondissipative state $\rho_{\rm nd}^{\rm ls}$ that has only two
occupied energy levels, $e_1$ and $e_4$, with probabilities
$p_{{\rm nd}1}^{\rm ls}=0.6$ and $p_{{\rm nd}4}^{\rm ls}=0.4$ (and
temperature $\rho_{\rm nd}^{\rm ls}=2.466\,u/\Boltz $); in fact
the system has no lower entropy states $\rho$ that commute with
$H$, have energy $2u/5$, and have zero occupation probabilities
$p_n$ for every $n>4$ \cite{PRE}. The dots in the middle represent
the nondissipative state $\rho_{\rm nd}^{\rm ft}$ which appears as
the false target state during the first part of the trajectory,
plotted at the instant in time when the entropy of the
time-varying trajectory is equal to the entropy of this
 distribution.

 It is interesting to observe from Figures 1 (bottom subfigure), and
 Figure 2 (second subfigure)
  that during the early
 part of the trajectory, $\tau_D$ almost exactly coincides with
 $\tau_\Petwo$ while in the late part it almost exactly coincides with
 $\tau_\Pethree$, and the switch occurs when the trajectory slows
 down in the neighborhood of the `false target' nondissipative
 state.

 In Figure 1, the second subfigure shows the time dependence of
 the dimensionless entropy $\mean{S}/\Boltz$; the third subfigure shows its rate of
 change (proportional to $\scov{M}{M}$) and compares it with $\scov{S}{S}$
 and $\scov{H}{H}/\theta^2$, to illustrate relation (\ref{covSM}); the
 fourth shows the time dependence of our generalized `nonequilibrium
 temperature' $\theta$ (properly nondimensionalized) and compares
 it with $\Delta_H/\Delta_S$ and $2\Delta_M\Delta_H/\scov{S}{S}$
 to illustrate relations (\ref{thetabound}) and
 (\ref{thetabound2}); the fifth subfigure shows the time
 dependence of $1/\tau_D$ (which here is proportional to the square
 root of the rate of entropy generation, third subfigure) and compares
 it with $1/\tau_K$ and $1/\tau_S$ to illustrate relations
 (\ref{deftauG}) and (\ref{tauSD}); the sixth subfigure shows
 $1/\tau_\Pen$ for each of the four occupation probabilities and
 compares them with  $1/\tau_D$ to illustrate relation
 (\ref{teuPen}), which for this particular trajectory has the
 feature we just discussed.

 In Figure 2, the second subfigure illustrates again relation
 (\ref{teuPen}) for each of the four observables
 $p_n=\mean{\Pen}$; the third subfigure illustrates the time-entropy uncertainty relation
 (\ref{genunc12}) for the same observables; the fourth illustrates
 inequality (\ref{dpndt}); the fifth illustrates relations (\ref{tauSD}) and
 (\ref{genunc5}).

 Similar remarks hold for Figures 3 and 4, where it is noteworthy
 that most qualitative features remain the same, except for the
 almost
 singular behavior near canonically distributed nondissipative states,
 where  $\Delta_M$ approaches zero and so does the dissipative time  $\tau$
 assumed in this case according to (\ref{pdotdefs2}). The approach
 to (partial) equilibrium in this case is not exponential in time as for
 $\tau =$ const. This puzzling behavior suggests that
  assumption (\ref{pdotdefs2}) may hardly be
 physically sensible. However, as already noted after (\ref{dissTEA}), it represents an
 extreme behavior, i.e., the minimum dissipative time functional $\tau(\rho)$
  by which observables
 which commute with $H$, like the occupations $\Pen$,  never violate the usual time-energy
 uncertainty relations $\tau_\Pen\Delta_H\ge \hbar/2$, even though
 their time dependence is not determined here by unitary dynamics but
 by purely dissipative dynamics.   These usual time-energy
 uncertainty relations, $\tau_\Pen\ge\tau_U$, are illustrated by the second subfigure of Fig. 4, because
  in this case $\tau_U=\tau_D$.

\section{\label{Conclusions}Conclusions}

The  Mandelstam-Tamm-Messiah time-energy uncertainty relation
    $\tau_{F}\Delta_H\ge \hbar/2$ provides a general lower bound
    to the characteristic times of change of all observables of a
    quantum system that can be expressed as linear functionals of
    the density operator $\rho$. This has been used to obtain
    estimates of rates of change and lifetimes of unstable states,
    without solving explicitly the time dependent evolution
    equation of the system. It may also be used as a general
    consistency check in measurements of time dependent phenomena.
    In this respect, the exact relation and inequalities
    (\ref{exactTE}) [that we derive for standard unitary dynamics based
    on the generalized Schr\"odinger inequality
    (\ref{Sinequality})] provide a more general and sharper chain
    of consistency checks than the usual time-energy uncertainty relation.

    The growing interest during the last several decades in
    quantum dynamical models of systems undergoing irreversible
    processes has been motivated by impressive technological
    advances in the manipulation of smaller and smaller systems,
    from the micrometer scale to the nanometer scale, and down to
    the single atom scale. The laws of thermodynamics, that fifty
    years ago were invariably understood as pertaining only to
    macroscopic phenomena, have gradually earned more attention and
    a central rule in studies of mesoscopic phenomena first, and of
    microscopic phenomena more recently. In this paper we do not address
     the controversial issues currently under discussion
    about interpretational matters, nor do we attempt a reconstruction and
    review of the different views,  detailed models and  pioneering contributions
    that propelled this fascinating advance of thermodynamics
    towards the realm of few particle and single particle systems.

    However, motivated by this context and background, we derive
    extensions of the usual time-energy uncertainty relations that
     extend their usefulness to studies of dissipative phenomena.
     We do this by focusing
 on a special but broad class of model evolution
    equations, that has been designed for the description of dissipative
    quantum phenomena and for satisfying a set of compatibility conditions
    with  general thermodynamic principles,
    and we derive in this framework various forms of
    time-energy and time-entropy
    uncertainty relations, and other interesting general inequalities, that
    should turn out to be  useful as additional
    consistency check in measurements of time-dependent dissipative phenomena.
    Finally, we illustrate
    and discuss some of
    these relations with numerical results obtained by integration of the nonlinear evolution
    equation introduced by this author for the description of
    steepest entropy ascent dynamics of an isolated system far
    from thermodynamic equilibrium.

\appendix
\section{\label{Appendix}Reasons for not assuming a
Kossakowski-Lindblad form of the evolution equation}

With various motivations, fundamental or phenomenological, a
number of different generalizations of quantum dynamics have been
proposed in which the evolution equation for the density operator
$\rho$ does not conserve the functional $-\Tr(\rho\ln\rho)$. In
particular, in theories of systems in contact with a heat bath, or
subsystems of composite systems which as a whole evolve unitarily,
a variety of successful model evolution equations for the reduced
density operator have the Kossakowski-Lindblad \cite{Lindblad}
form
\begin{equation}
\label{Lindblad}
\ddt{\rho}=-\frac{i}{\hbar}[H,\rho]+\frac{1}{2}\sum_j\left(2
V_j^\dagger\rho  V_j -\{ V_j^\dagger V_j,\rho\}\right) \ ,
\end{equation}
where the $V_j$'s are operators on $\Hil$ (each term within the
summation, often written in the alternative form $[V_j,\rho
V_j^\dagger]+ [V_j\rho, V_j^\dagger]$, is obviously traceless).
Evolution equations of this form are linear in the density
operator $\rho$ and preserve its hermiticity, nonnegativity and
trace \cite{Simmons2}.

For example, in a number of successful models of dissipative
quantum dynamics of open subsystems,  operators $V_j$ are in
general interpreted as creation and annihilation, or transition
operators. For example, by choosing $V_j=c_{rs}|r\rangle\langle
s|$,  where $ c_{rs} $ are complex scalars and $ | s\rangle $
eigenvectors of the Hamiltonian operator $H$, and defining the
transition probabilities $w_{r s}= c_{rs} c_{rs}^*  $, equation
(\ref{Lindblad}) becomes
\begin{equation}
\label{Pauli1} \ddt{\rho} = - \, \fr{i}{\hbar} [ H , \, \rho ] +
\sum_{r s} w_{r s} \left( | s \rangle \langle\rho \rangle\langle s
| - \, \fr{1}{2} \{ | s \rangle\langle s | \, , \rho \} \right) \,
,
\end{equation}
or, equivalently, for the $ nm $-th matrix element of $\rho$ in
the $H$ representation,
\begin{eqnarray}\label{Pauli2}&& \ddt{\rho_{nm}} =
 -\fr{i}{\hbar} \rho_{nm} ( E_n - E_m )\nonumber\\ &&+ \delta_{nm}
 \sum_r w_{n r} \rho_{r
r} - \rho_{nm} \fr{1}{2} \sum_r ( w_{r n} + w_{r m} ) \ ,
\end{eqnarray}
which, for the occupation probabilities $p_n=\rho_{nn}$, is the
Pauli master equation
\begin{equation}\label{Pauli3} \ddt{ p_n} = \sum_r w_{n r} p_r -
p_n \sum_r w_{r n} \  .
\end{equation}

In this paper, we consider a  class of model evolution equations
applicable not only to  open systems but also to
 closed isolated systems, capable of describing, simultaneously with the usual
Hamiltonian unitary evolution, the natural tendency of any initial
nonequilibrium state to relax towards canonical or
partially-canonical thermodynamic equilibrium, i.e., capable of
describing the irreversible tendency to evolve towards the highest
entropy state compatible with the instantaneous mean values of the
energy, the other constants of the motion, and possibly other
constraints. To avoid the severe restrictions imposed by the
linearity of the evolution equation, we open our attention to
evolution equations nonlinear in the density operator $\rho$.
Therefore, it may at first appear natural to maintain the
Kossakowski-Lindblad form (\ref{Lindblad}) and simply assume that
operators $V_j$ are functions of $\rho$. This is true only in part
for the evolution equation (\ref{rhodotHM}) that we assume.
Indeed, our hermitian operator $\Delta M(\rho)/\Boltz\tau(\rho)$
can always be written as $-\sum_j V_j^\dagger(\rho) V_j(\rho)$ and
therefore our anticommutator term may be viewed as a
generalization of the corresponding term in (\ref{Lindblad}).

However, in our equations (\ref{rhodot}) and  (\ref{rhodotHM}) we
suppress the term corresponding to  $\sum_j V_j^\dagger\rho V_j$
in (\ref{Lindblad}). The reason for this suppression is the
following. Due to the terms $V_j^\dagger\rho V_j$, whenever the
state operator $\rho$ is singular, i.e., it has one or more zero
eigenvalues, Eq. (\ref{Lindblad})  implies that these zero
eigenvalues may change at a finite rate. This can be seen clearly
from (\ref{Pauli3}) by which ${\rm d}p_n/{\rm d}t$ is  finite
whenever there is a nonzero transition probability $w_{nr}$ from
some other populated level ($p_r\ne 0$), regardless of whether
$p_n$ is zero or not. When this occurs, for one instant in time
the rate of entropy change is infinite, as seen clearly from the
expression of the rate of entropy change implied by
(\ref{Lindblad}),
\begin{eqnarray}\label{PauliS1} \ddt{\mean{S}} &=& \Boltz \sum_j
\Tr ( V_j^{\dagger}
 V_j \rho \ln
\rho - V_j^{\dagger} \rho V_j \ln \rho )\nonumber\\ &=& \Boltz
\sum_{j r n}
 ( V_j )^*_{n r}  ( V_j )_{n r} ( \rho_r - \rho_n ) \ln \rho_r \ ,
\end{eqnarray}
where $\rho_r$ denotes the $r$-th eigenvalue of $\rho$ and $( V_j
)_{n r}$ the matrix elements of $V_j$ in the $\rho$
representation.

We may argue that an infinite rate of entropy change  can  be
tolerated, because it would last only for one instant in time. But
the fact that zero eigenvalues of $\rho$ in general could not
survive, i.e., would not remain zero (or close to zero) for longer
than one instant in time, is an unphysical feature, at least
because it is in contrast with a wealth of successful models of
physical systems in which great simplification is achieved  by
limiting our attention to a restricted subset of relevant
eigenstates (forming a subspace of $\Hil$ that we call the
effective Hilbert space of the system \cite{MPLA}). Such common
practice models yield extremely good results, that being
reproducible, ought to be relatively robust with respect to
including in the model other less relevant eigenstates. In fact,
such added
 eigenstates, when initially unpopulated, are irrelevant if
 they remain unpopulated
(or very little populated) for long times, so that neglecting
their existence introduces very little error. The terms
$V_j^\dagger\rho V_j$, instead, would rapidly populate such
irrelevant unpopulated eigenstates and void the validity of our so
successful simple models, unless we deliberately  overlook this
instability problem by highly ad-hoc assumption, e.g., by forcing
the $V_j$'s to be such that $( V_j )_{n r}=0$ whenever either
$\rho_n=0$ or $\rho_r=0$, in which case, however, we can no longer
claim true linearity with respect to $\rho$.

To avoid the unphysical implications of this seldom recognized
\cite{MPLA,Beretta} problem of linear evolution equations of form
(\ref{Lindblad}), we consider in this paper only equations of form
(\ref{rhodotHM}). We do not exclude that it may be interesting to
investigate also the behavior of equations that include  nonlinear
terms of the form $V_j^\dagger(\rho)\, \rho\, V_j(\rho)$. However,
at least when the system is strictly isolated, the
operator-functions $V_j(\rho)$ should be such that $( V_j(\rho)
)_{n r}=0$ whenever either $\rho_n=0$ or $\rho_r=0$.

Another important general physical reason why we exclude terms
that generate nonzero rates of change of zero eigenvalues of
$\rho$, is that if such terms are construed so as to conserve
positivity in forward time, in general they cannot maintain
positivity in backward time. The view implicitly assumed when Eq.
(\ref{Lindblad}) is adopted, is that the model is ``mathematically
irreversible'' (a distinguishing feature if not a starting point
of the theory of completely positive linear dynamical semigroups
on which it is based), in the sense that neither uniqueness of
solutions in forward time nor existence in backward time are
required (and granted). Such mathematical irreversibility of the
initial value problem, is often accepted, presented and justified
as a natural counterpart of physical irreversibility. However, it
is more related to the principle of causality than to physical
irreversibility. The strongest form of the non-relativistic
principle of causality, a keystone of traditional physical
thought, requires that future states of a system should unfold
deterministically from initial states along smooth unique
trajectories in state domain defined for all times (future as well
as past). Accepting mathematical irreversibility of the model
dynamics, implies giving up such causality requirement. But it
 is not strictly necessary  to describe physical irreversibility,
 at least not if we are willing  to give up linearity instead. The
proof of this statement is our Eq. (\ref{rhodotHM}) which,
together with the additional assumptions made in Section
\ref{Example} to describe relaxation within an isolated system, is
mathematically reversible, in the sense that it features existence
and uniqueness of well-defined solutions both in forward and
backward time, and yet it does describe physically irreversible
time evolutions, in the sense that the physical property described
by the entropy functional $-\Boltz\Tr(\rho\ln\rho)$ is a strictly
increasing function of time for all states except  the very
restricted subset defined by Eq. (\ref{rhonondiss}), where it is
time invariant.

\begin{figure}
\begin{center}
\includegraphics[width=0.40\textwidth]{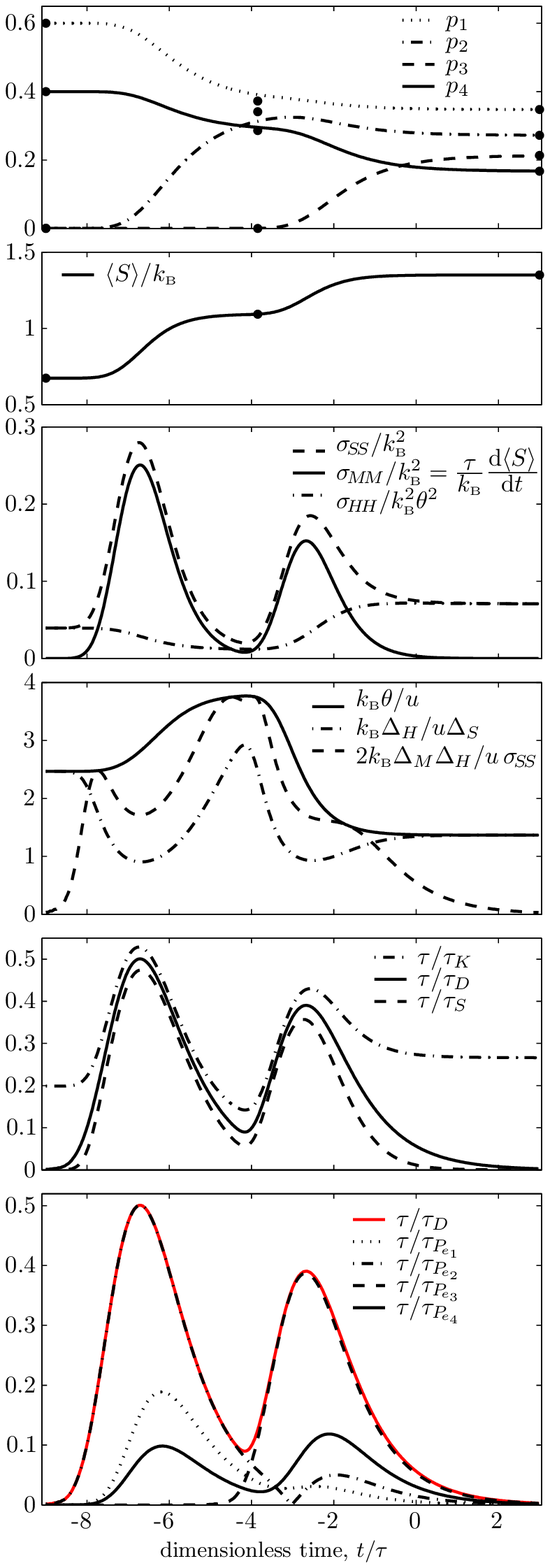}
 \caption{\label{Figure1}(Color online) See text; $\tau$ assumed constant.}
\end{center}
\end{figure}

\begin{figure}
\begin{center}
\includegraphics[width=0.40\textwidth]{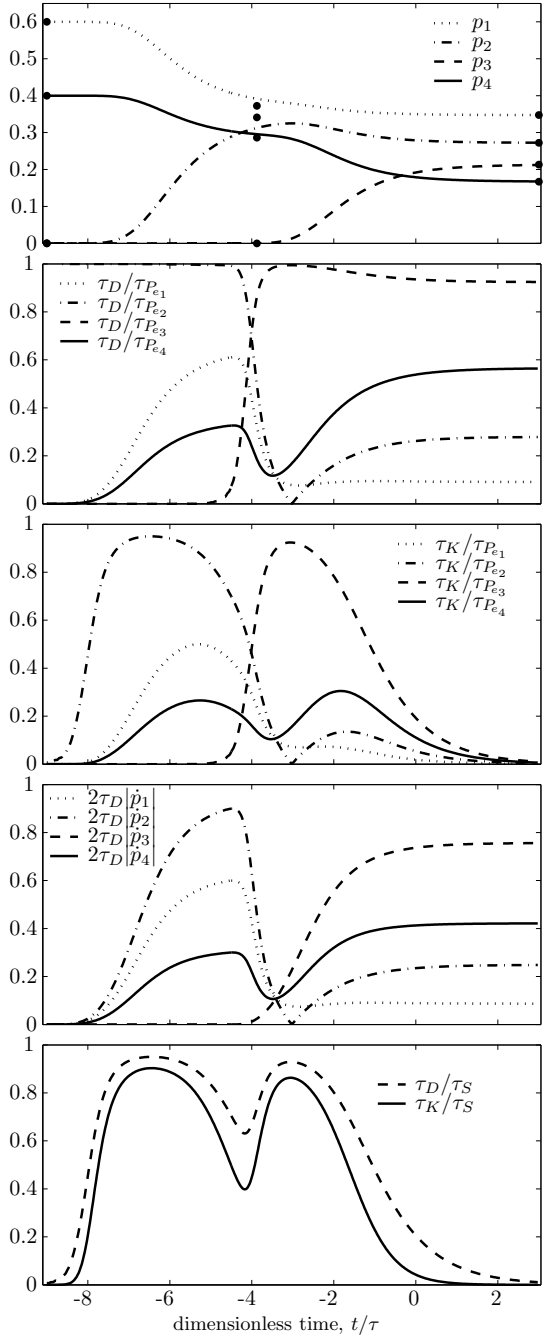}
\caption{\label{Figure2}See text; $\tau$ assumed constant.}
\end{center}
\end{figure}

\begin{figure}
\begin{center}
\includegraphics[width=0.40\textwidth]{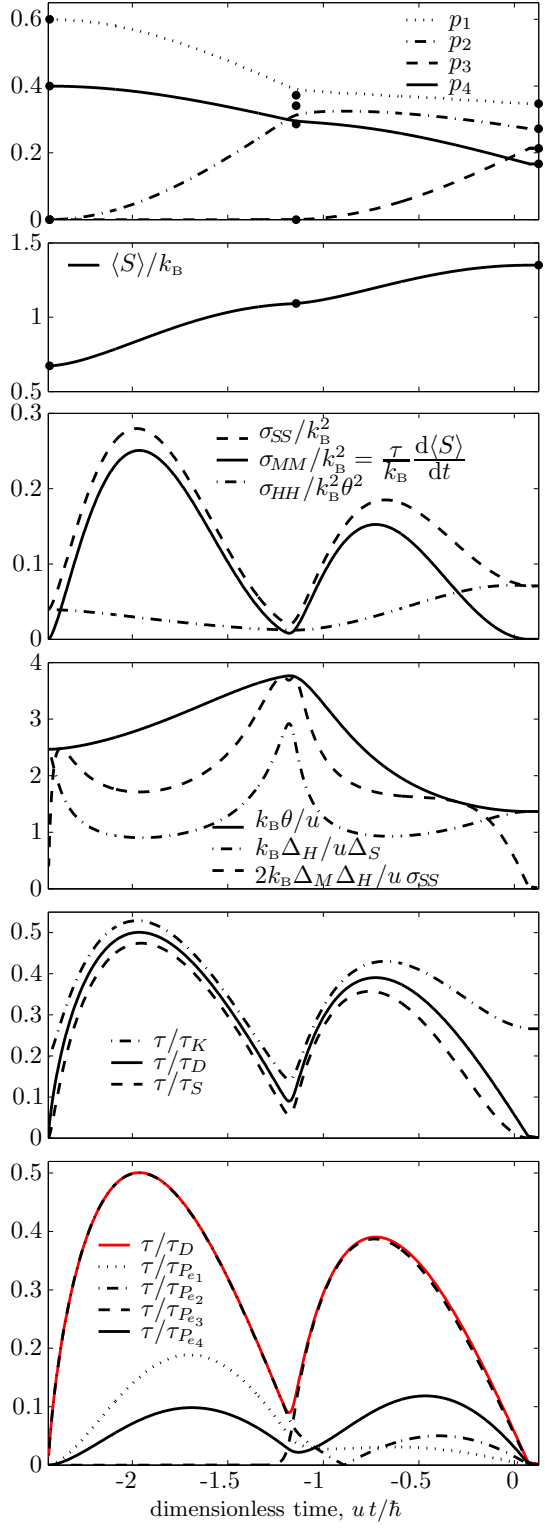}
\caption{\label{Figure3}(Color online) See text; $\tau$ assumed
according to (\ref{pdotdefs2}).}
\end{center}
\end{figure}

\begin{figure}
\begin{center}
\includegraphics[width=0.40\textwidth]{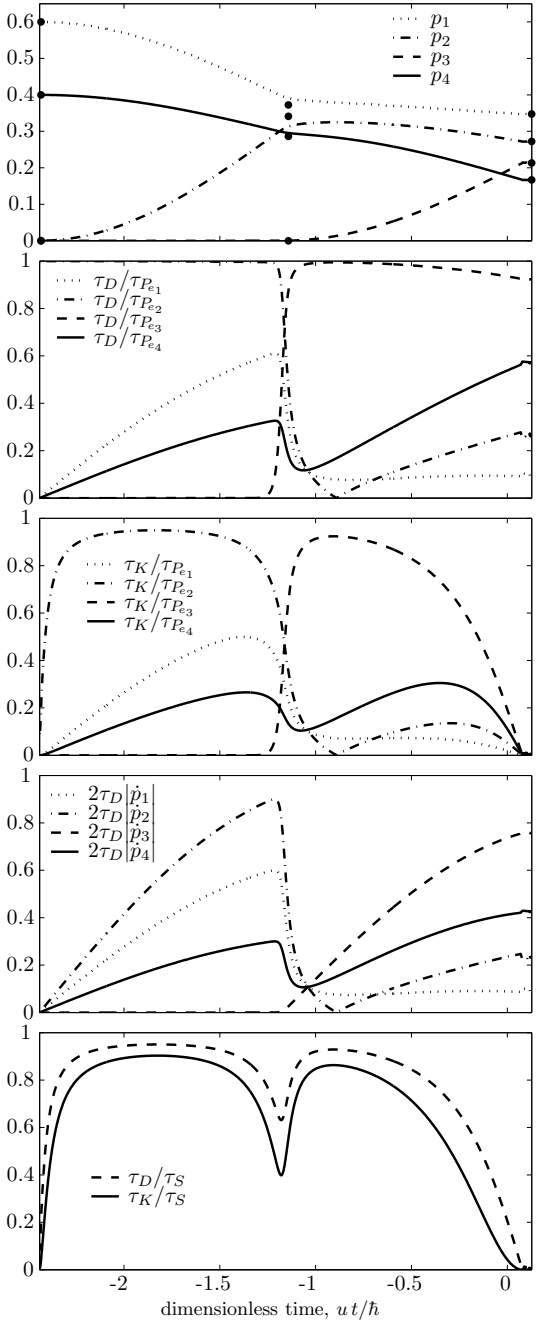}
\caption{\label{Figure4}See text; $\tau$ assumed according to
(\ref{pdotdefs2}).}
\end{center}
\end{figure}



\begin{thebibliography}{33}
\bibitem{operatorB}Operator $S$, introduced in \cite{Beretta},
 is always well defined for any $\rho\ge\rho^2$. It is
the null operator when $\rho^2=\rho$. In general, $\sq S=
-\Boltz\sq\ln\rho$.
\bibitem{proofmaximal}Operator $\sq E$ is a `vector' in
$\Ell(\Hil)$ and determines through its scalar product with $\sq
F$ and $\sq S$ [Eqs.\ (\ref{rateF}) and (\ref{rateS})] the rates
of change of $\Tr(\rho F)$ and $\Tr(\rho S)$, respectively. From
(\ref{rateS}) and the Schwarz inequality $(\sq S|\sq E)^2\le (\sq
S|\sq S)(\sq E|\sq E)$, we see that for a given $\rho$, among all
 vectors $\sq E$ with  given norm $(\sq E|\sq
E)=c(\rho)$, the one  maximizing $(\sq S|\sq E)$ has the same
direction as $\sq S$. In general, along such direction $\Tr(\rho
H)$ and $\Tr(\rho\N )$ are  not conserved because $\sq S$ is not
always  orthogonal to $\sq H$ and $\sq \N$. Instead, dynamics
along the direction of steepest entropy ascent compatible with
such conservation requirements, as first postulated and formulated
in \cite{Beretta}, obtains when $\sq E$ has the direction of the
component of $\sq S$ orthogonal to $\sq H$ and $\sq \N$. This is
precisely how $\Delta M(\rho)$ is defined through Eqs.\
(\ref{Msteepest})-(\ref{constraint2}). See also Ref.\
\citen{ArXiv1}.
\end{thebibliography}
\end{document}